\documentclass[prd,twocolumn,nofootinbib,superscriptaddress,eqsecnum,longbibliography]{revtex4-1}

\usepackage[utf8]{inputenc}
\usepackage{mathtools}
\usepackage{graphics}
\usepackage{graphicx}
\usepackage{float}
\usepackage{dcolumn}
\usepackage{bm}
\usepackage{dsfont} 
\usepackage{amsmath}
\usepackage{amssymb}
\usepackage{hyperref}
\usepackage{tabularx}
\usepackage{epstopdf}
\usepackage{tensor}
\usepackage{mathrsfs}
\usepackage{breqn}

\usepackage[dvipsnames]{xcolor}
\usepackage{mathrsfs}

\usepackage[normalem]{ulem}

\hypersetup{
    colorlinks=true,
    linkcolor=NavyBlue,
    filecolor=Magenta,      
    urlcolor=NavyBlue,
    citecolor=NavyBlue
}

\newcommand{\upd}[1]{\textcolor{black}{{#1}}}

\newcommand{\GR}{{\mbox{\tiny GR}}}

\newcommand\be{\begin{equation}}
\newcommand\ba{\begin{eqnarray}}
\newcommand\ee{\end{equation}}
\newcommand\ea{\end{eqnarray}}
\newcommand\bw{\begin{widetext}}
\newcommand\ew{\end{widetext}}
\newcommand\dd{{\rm d}}

\newcommand\nn{\nonumber}

\newcommand\dfz{f^{\prime}_0}
\newcommand\dfzs{f_0^{\prime 2}}
\newcommand\ddfz{f^{\prime \prime}_0}

\AtBeginDocument{%
    \newwrite\bibnotes
    \def\bibnotesext{Notes.bib}
    \immediate\openout\bibnotes=\jobname\bibnotesext
    \immediate\write\bibnotes{@CONTROL{REVTEX41Control}}
    \immediate\write\bibnotes{@CONTROL{%
    apsrev41Control,author="08",editor="1",pages="1",title="0",year="1"}}
    \if@filesw
    \immediate\write\@auxout{\string\citation{apsrev41Control}}%
    \fi
}

\newcommand\rpk[1]{r_{\rm m}^{#1}}
\newcommand\cG{\mathscr{G}}

\newcommand{\UVA}{Department of Physics, University of Virginia, Charlottesville, Virginia 22904-4714, USA}
\newcommand{\UIUC}{Illinois  Center  for  Advanced  Studies  of  the  Universe \& 
Department of Physics, University of Illinois at Urbana-Champaign, Urbana, Illinois 61801, USA}
\newcommand{\UM}{Departamento de F\'isica, Universidad de Murcia, Murcia, E-30100, Spain}
\newcommand{\UT}{Theoretical Astrophysics, University of T\"ubingen, Auf der Morgenstelle 10, T\"ubingen, D-72076, Germany}
\newcommand{\AEI}{Max Planck Institute for Gravitational Physics (Albert Einstein Institute), Am M\"uhlenberg 1, Potsdam 14476, Germany}

\begin{document}

\title{Eikonal quasinormal modes of black holes beyond general relativity III: \\ scalar Gauss-Bonnet gravity}

\author{Albert Bryant}
\affiliation{\UVA}

\author{Hector O. Silva}
\affiliation{\AEI}
\affiliation{\UIUC}

\author{Kent Yagi}
\affiliation{\UVA}

\author{Kostas Glampedakis}
\affiliation{\UM}
\affiliation{\UT}

\date{\today}

\begin{abstract}
In a recent series of papers we have shown how the eikonal/geometrical optics approximation can be used to calculate 
analytically the fundamental quasinormal mode frequencies associated with coupled systems of wave equations, which 
arise, for instance, in the study of perturbations of black holes in gravity theories beyond General Relativity.
As a continuation to this series, we here focus on the quasinormal modes of nonrotating black holes in scalar Gauss-Bonnet gravity
assuming a small-coupling expansion.
We show that the axial perturbations are purely tensorial and are described by a modified Regge-Wheeler equation, while 
the polar perturbations are of mixed scalar-tensor character and are described by a system of two coupled wave equations.
When applied to these equations, the eikonal machinery leads to axial
quasinormal modes that deviate from the general relativistic results 
at quadratic order 
in the Gauss-Bonnet coupling constant.
We show that this result is in agreement with an analysis of unstable circular null orbits
around black holes in this theory, allowing us to establish the geometrical optics--null geodesic
correspondence for the axial quasinormal modes.
For the polar quasinormal modes the small-coupling approximation forces us to consider 
the ordering between eikonal and small-coupling perturbative parameters; one of which we
show, by explicit comparison against numerical data, yields the correct identification of 
the quasinormal modes of the scalar-tensor coupled system of wave equations.
These corrections lift the general relativistic degeneracy 
between scalar and tensorial eikonal quasinormal modes at quadratic order in Gauss-Bonnet coupling 
in a way reminiscent of the Zeeman effect.
In general, our analytic, eikonal quasinormal mode frequencies (normalized to the General Relativity ones) agree with numerical results with an error of $\mathcal{O}(10\%)$ in the regime of small coupling constant.
Finally, we find that the analytical expressions for the quasinormal modes are common 
to a broad class of scalar-Gauss-Bonnet theories to leading eikonal order, showing a degeneracy between 
the quasinormal modes of nonrotating black holes in particular scalar-Gauss-Bonnet theories 
in the geometrical optics limit.
\end{abstract}

\maketitle

\section{Introduction} 
\label{sec:intro}

The first direct observations of gravitational waves (GWs) by the LIGO/Virgo Collaborations marked the dawn of gravitational-wave astronomy~\cite{GW150914,Abbott2017,Monitor:2017mdv,GW_Catalogue,Abbott:2020niy}. 
These GW events allow us to probe gravity in the strong, dynamical and nonlinear regime~\cite{Berti:2015itd,Yunes_ModifiedPhysics} and
to compare the predictions of general relativity (GR), and modifications thereof, in such extreme environments as done e.g.~in~\cite{Yunes_ModifiedPhysics,Abbott_IMRcon2,Abbott_IMRcon,Abbott:2018lct,Berti:2018cxi,Nair:2019iur,Perkins:2021mhb}.
An example is the inspiral-merger-ringdown consistency test in a coalescing binary system~\cite{Ghosh_IMRcon,Ghosh_IMRcon2};
this is a consistency check between the independent measurements of the remnant black hole's mass 
and spin from the inspiral and merger-ringdown phases, assuming GR is correct. 
Such consistency tests can be applied beyond the realm of GR to constrain specific theories~\cite{Carson:2020cqb} and parametrized deformed-Kerr spacetimes~\cite{Carson:2020iik}.

A similar suit of consistency tests can be performed with the ringdown signal alone, with the aim of 
probing the no-hair property of black holes~\cite{Berti:2018vdi}. 
In this approach, 
usually termed ``black hole spectroscopy'', a measurement of the fundamental quasinormal mode (QNM) frequency and damping time allows the extraction of the remnant's mass and spin under the assumption that the object is a garden-variety Kerr black hole. A much more powerful test -- that of the Kerr hypothesis itself --
can be performed if additional QNM frequencies can be observed in the data stream.
Indeed, the very first event GW150914 has been analysed in this fashion using overtones~\cite{Isi:2019aib}. 
A more traditional approach is to use waveforms of the same overtone but at different harmonics~\cite{Berti:2005ys}, which has successfully been applied recently to e.g. GW190521~\cite{Capano:2021etf}.
This ``spectroscopic approach'' can be applied to test gravity both in theory-specific~\cite{Yunes:2007ss,Ferrari:2000ep,Molina:2010fb,Blazquez-Salcedo:2016enn,Blazquez-Salcedo:2020caw,Blazquez-Salcedo:2020rhf,Blazquez-Salcedo:2018,Blazquez_Salcedo:2017,Brito:2018hjh,Bao:2019kgt,Moulin:2019ekf,Tattersall:2018nve,Tattersall:2019pvx,Cano:2020cao,Wagle:2021tam,Pierini:2021jxd,Blazquez-Salcedo:2020caw} and model-independent~\cite{Glampedakis:2017dvb,Brito:2018rfr,Cardoso:2019mqo,McManus:2019ulj,Maselli:2019mjd,Kimura:2020mrh,Abbott:2020jks,Carullo:2021dui,Ghosh:2021mrv} frameworks.

This paper makes a contribution to the former category by computing QNMs of black holes in scalar Gauss-Bonnet gravity with the help of the eikonal approximation. The action of this theory features a scalar field non-minimally coupled to the Gauss-Bonnet invariant (which itself is quadratic in curvature)~\cite{Antoniou:2017acq,Antoniou:2017hxj}. The precise functional form of this coupling gives rise to 
different sub-theories of gravity. For example, an exponential scalar field coupling can be identified 
as the Einstein-dilaton Gauss-Bonnet (EdGB) gravity motivated by string theory~\cite{Gross:1986mw,Metsaev:1987zx,Kanti:1995vq,Maeda:2009uy}. 
On the other hand, a linear coupling leads to a shift-symmetric theory~\cite{Yunes:2011we,Yagi:2011xp,Sotiriou:2013qea,Sotiriou:2014pfa,Yagi:2015oca} 
while theories with a quadratic coupling lead to spontaneously scalarized black holes~\cite{Silva:2017uqg,Dima:2020yac,Berti:2020kgk,Silva:2020omi} 
(this effect can also occur with other coupling functions and scalar field self-interactions, see e.g.~\cite{Doneva:2017bvd,Doneva:2017duq,Silva:2018qhn,Minamitsuji:2018xde,Macedo:2019sem,Andreou:2019ikc,Herdeiro:2020wei} for details).  

In this paper we consider the broader scalar Gauss-Bonnet gravity theory
and study the QNMs of its spherically symmetric, nonrotating black holes using the eikonal approximation.
We achieve this by first solving the linearized field equations describing combined
scalar-tensor perturbations of black holes. The final distilled wave
equations for the decoupled polar and axial degrees of freedom can be cast 
in a Schr\"odinger-like form. 
These equations are subsequently solved using the eikonal techniques we developed 
in~\cite{Glampedakis:2019dqh,Silva:2019scu} in the context of non-GR theories. 
The end result (summarized in Sec.~\ref{sec:summary}) is a set of analytic eikonal formulae for the fundamental QNM's 
frequency and damping time. In order to gauge the accuracy of our formulation
we consider the particular example of EdGB gravity and compare our results against the
numerical QNM data computed in~\cite{Blazquez-Salcedo:2016enn}. 

\begin{figure}
\includegraphics[width=8.5cm]{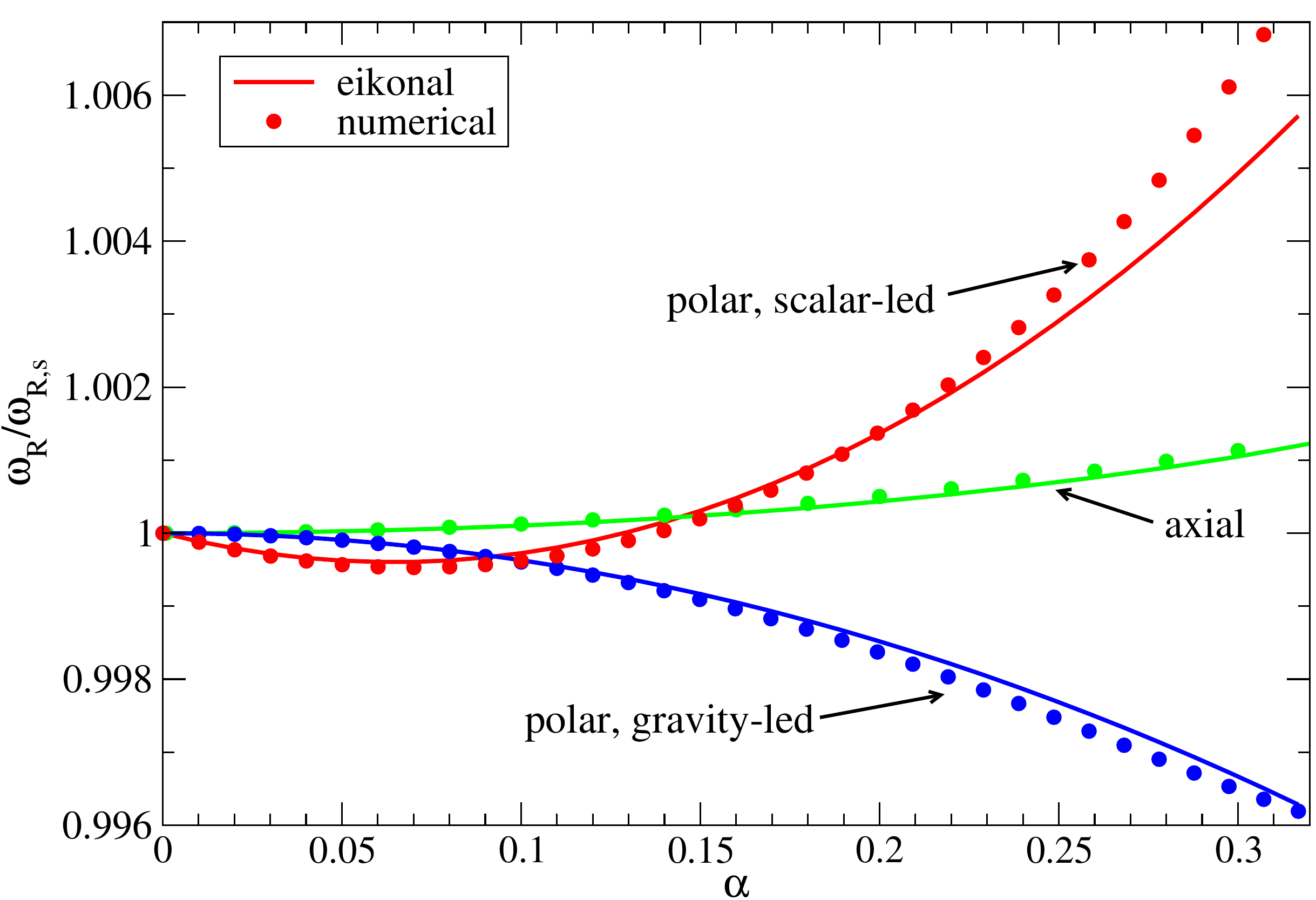}
\caption{\label{fig:summary}
Summary plot comparing eikonal and numerical real QNM frequencies (normalized by the GR Schwarzschild value) for $\ell=2$ in EdGB gravity as a function of the coupling parameter $\alpha$ (in units of GR black hole mass squared). The eikonal result is computed within the small coupling approximation (valid to $\mathcal{O}(\alpha^2)$) and thus becomes more inaccurate for larger $\alpha$.
}
\end{figure}

Figure~\ref{fig:summary} compares the (normalized) real eikonal QNM frequencies for the 
$\ell = 2$ harmonic
in EdGB gravity against \upd{the} numerical results 
of Ref.~\cite{Blazquez-Salcedo:2016enn}
as a function of the coupling constant $\alpha$ in the theory. Notice that the analytic, eikonal results match nicely with the numerical ones in the small $\alpha$ regime. We found that the former is accurate with an error of $\sim 10\%$. The eikonal results become less accurate for larger $\alpha$ as they are derived within the small coupling approximation.     

The rest of the paper is organized as follows. 
In Sec.~\ref{sec:sgb_overview} we review the basics of scalar Gauss-Bonnet gravity
and show the nonrotating black hole solution in this theory, which is then perturbed in Sec.~\ref{sec:BH_pert}.
Having derived the master equations governing gravito-scalar 
perturbations of black holes in this theory, we examine them under the lens of the eikonal limit in Sec.~\ref{sec:eikonal_qnm}.
We compare our eikonal results with the numerical ones in Sec.~\ref{sec:comparison} and summarize the final eikonal expressions in Sec.~\ref{sec:summary}.
We present our conclusions in Sec.~\ref{sec:conclusions} and give directions for potential future work.
We work with geometrical units $c=G=1$. Throughout the paper a prime denotes 
a derivative with respect to a function's argument.

\section{Scalar Gauss-Bonnet Gravity}
\label{sec:sgb_overview}

We begin by reviewing the theory and a nonrotating black hole spacetime in scalar-Gauss-Bonnet gravity.

\subsection{Theory}

Our starting point is the action for scalar Gauss-Bonnet gravity~\cite{Antoniou:2017hxj}:
\begin{align}
\label{eq:action}
S &= \frac{1}{16\pi} \int \dd^4x \sqrt{-g} \, 
\left[
R-\frac{1}{2}\partial_a\phi\,\partial^a\phi
+
\alpha f(\phi) \cG
\right]
+ S_m 
\,,
\end{align}
where
\begin{equation}
\label{eq:GB}
\cG = R_{abcd}R^{abcd} - 4R_{ab}R^{ab} + R^2\,,
\end{equation}
is the Gauss-Bonnet topological term and $S_m$ is the matter part of the action. Different choices of the arbitrary scalar field function $f(\phi)$ correspond to different flavors of scalar Gauss-Bonnet theory. For example, the popular choice $f(\phi) \propto \exp(\gamma \phi)$, where $\gamma$ is a constant, corresponds to EdGB gravity that arises in the low-energy limit of string theories~\cite{Gross:1986mw,Metsaev:1987zx,Kanti:1995vq,Maeda:2009uy}; $f(\phi) \propto \phi$ corresponds to shift-symmetric scalar Gauss-Bonnet theory~\cite{Yunes:2011we,Yagi:2011xp,Sotiriou:2013qea,Sotiriou:2014pfa,Yagi:2015oca}; 
the class of theories with $f(\phi) \propto\phi^2$~\cite{Silva:2017uqg}
and $f(\phi) \propto \exp(\gamma \phi^2)$~\cite{Doneva:2017bvd} has been recently considered within the context of spontaneous scalarizations of black holes and neutron stars.

A standard variation of the action
returns the field equations
\begin{align}
\label{eq:scalar}
\Box\phi &= \alpha f'(\phi) \cG\,,  \\
\label{eq:gravity}
G_{ab} &= \frac{1}{2}\partial_a\phi\,\partial_b\phi-\frac{1}{4}g_{ab}\partial_c\phi\, \partial^c\phi-\alpha\mathcal{K}_{ab}+8\pi T_{ab}\,,
\end{align}
where $G_{ab}$ is the usual Einstein tensor, $T_{ab}$ is the matter stress-energy tensor, and
\be
\label{eq:Ktensor}
\mathcal{K}_{ab}=(g_{ac}g_{bd}+g_{ad}g_{bc}) \, 
\epsilon^{idjk}\nabla_l \,
[{}^{\ast}{R}^{cl}_{\hspace{2mm}jk}\partial_if(\phi)]\,,
\ee
which arises from the Gauss-Bonnet term to the action, where
$\epsilon^{abcd}$ is the Levi-Civita pseudotensor and
 ${}^{\ast}{R}^{ab}_{\hspace{3mm}cd}=\epsilon^{abij} R_{ijcd}$ 
is the dual to the Riemann tensor.

\subsection{The background black hole spacetime}
\label{sec:BH_bg}
Nonrotating black holes in 
scalar Gauss-Bonnet gravity can be described by the \upd{static and} spherically symmetric line element
\begin{equation}
g_{ab}^0 \dd x^a \dd x^b 
= -A(r) \dd t^2 + B(r)^{-1} \dd r^2 + r^2 \dd \Omega^2\,,
\label{eq:line_element}
\end{equation}
where $\dd \Omega^2$ is the unit two-sphere line element.
Hereafter we work with dimensionless quantities, i.e.,
\begin{equation}
r/M \to r,~ x/M \to x, ~\alpha/M^2 \to \alpha
\,,
\end{equation}
where $M$ is the black hole's mass in GR.
The metric functions $A$ and $B$ were obtained in the past, both in shift-symmetric 
and dilatonic flavors of scalar-Gauss-Bonnet gravity, working either perturbatively in a small $\alpha$ expansion around a seed Schwarzschild background black hole (see e.g.~\cite{Mignemi:1992nt,Sotiriou:2013qea,Sotiriou:2014pfa,Delgado:2020rev}) or by directly integrating the field equations numerically (see e.g.~\cite{Kanti:1995vq,Alexeev:1996vs,Maeda:2009uy,Antoniou:2017hxj,Silva:2017uqg,Doneva:2017bvd}).

Here, $f$ is kept arbitrary for generality, but we do adopt a small coupling approximation ($\alpha \ll 1$) as done in~\cite{Julie:2019sab}.
In our coordinate system the background metric functions $A$, $B$ and scalar field $\phi_0$ can be written as
\bw
\begin{align}
A &= 1-\frac{2}{r} - \frac{\alpha^2 \dfzs}{r} 
\left(
\frac{49}{40} 
- \frac{1}{3 r^2}
- \frac{26}{3 r^3}
- \frac{22}{5 r^4}
- \frac{32}{5 r^5}
+ \frac{80}{3 r^6}
\right)
\,, 
\label{eq:metric_A}
\\
B &= 1-\frac{2}{r} - \frac{\alpha^2 \dfzs}{r} 
\left(
\frac{49}{40}
- \frac{1}{r}
- \frac{1}{r^2}
- \frac{52}{3 r^3}
- \frac{2}{r^4}
- \frac{16}{5 r^5}
+ \frac{368}{3 r^6}
\right)\,,
\label{eq:metric_B}
\\
\phi_0 &= \frac{2 \alpha \dfz}{r}
\left(
1 
+ \frac{1}{r} 
+ \frac{4}{3 r^2}
\right)
+ \frac{\alpha^2 \dfz \ddfz}{r}
\left(
\frac{73}{30}
+ \frac{73}{30 r}
+ \frac{146}{45 r^2}
+ \frac{73}{15 r^3}
+ \frac{224}{75 r^4}
+ \frac{16}{9 r^5}
\right)\,.
\label{eq:phi}
\end{align}
\ew
Here we used the shorthand notations $f'(0) = \dfz$ and $f''(0) = \ddfz$.
This solution represents a deformed, scalar hair-endowed Schwarzschild black hole, with deformations controlled by the parameter $\alpha$.
From Eq.~\eqref{eq:metric_A}, one finds that the \upd{Arnowitt-Deser-Misner (ADM)} mass $M_*$ of the black hole acquires an ${\cal O}(\alpha^2)$ correction as
\begin{equation}
\label{eq:ADM}
M_* = M\left(1 + \frac{49}{80} \alpha^2 \dfzs\right)\,.
\end{equation}

\section{Black Hole Perturbations}
\label{sec:BH_pert}

Going beyond the background spacetime, we now analyse its stability by studying linear perturbations.
We write the perturbed metric and scalar field as
\begin{equation}
\label{eq:metric_pert2}
g_{ab}=g_{ab}^0 + \bar\epsilon \, h_{ab}\,, 
\quad
\phi=\phi_0 + \bar\epsilon \, \delta\phi\,,
\end{equation}
\upd{where $\bar{\epsilon}$ is a bookkeeping parameter while} $g^0_{ab}$ and $\phi_0$ are given by 
Eqs.~\eqref{eq:line_element},~\eqref{eq:metric_A},~\eqref{eq:metric_B}, and \eqref{eq:phi}.

Following standard techniques of black hole perturbation theory in GR~\upd{\cite{Regge:1957td,Zerilli:1970se}}, we expand the metric/scalar field perturbations into appropriate tensor/scalar harmonics basis.
We work
to linear order in $\bar\epsilon$ and after imposing the Regge-Wheeler gauge, the field equations~\eqref{eq:scalar}--\eqref{eq:gravity} lead to a set of decouple equations for the axial and polar sectors of the perturbations. We discuss these separately in the following sections.

\subsection{Axial perturbations}
\label{sec:axial_pert}
We begin by considering axial perturbations, which
are decoupled from the scalar perturbations.
We follow the notation of Bl\'{a}zquez-Salcedo et al.~\cite{Blazquez-Salcedo:2016enn} where the axial perturbed metric is written as

\begin{gather}
\label{eq:metric_pert_YAX}
h_{ab}=\begin{pmatrix} 0 && 0 && 0 &&  \bar  h_0\sin\theta\partial_\theta\\
0 && 0 && 0 && \bar h_1 \sin\theta \partial_\theta\\
0 && 0 && 0 && 0\\
 \bar h_0\sin\theta\partial_\theta && \bar h_1\sin\theta \partial_\theta && 0 && 0
\end{pmatrix} Y_{\ell m}\,,
\end{gather}
where $Y_{\ell m}(\theta,\varphi)$ are the (scalar) spherical harmonics while $\bar h_0$ and $\bar h_1$ are functions of $t$ and $r$ only. We can further Fourier transform these functions as
\begin{equation}
\label{eq:Fourier}
\bar X(t,r) = \frac{1}{\sqrt{2\pi}} \int \upd{\dd}\omega X(r) e^{-i\omega t}\,,
\end{equation} 
with $X = (\bar h_0,\bar h_1)$.

Inserting Eqs.~\eqref{eq:metric_pert_YAX} and~\eqref{eq:Fourier} into the field equations, we find two non-trivial equations for the
axial gravitational perturbations. 
These equations, in the particular case of EdGB gravity with $f=e^\phi/4$, can be found in Appendix B of~\cite{Blazquez-Salcedo:2016enn}.
These perturbed field equations can be combined into a single
equation for $\bar h_1$ and its radial derivatives.
We can further make a field redefinition as
\begin{equation}
\label{eq:field_redefAX}
Q =c \bar h_1\,,
\end{equation}
where $c(r)$ is found by requiring that the coefficient of the friction term $Q'(r)$ vanishes (the expression can be found in Appendix~\ref{app:supplemental} and in the supplemental Mathematica notebook~\cite{mathematica}).
Then, we obtain a master equation for the axial perturbation, namely
\begin{equation}
\label{eq:masterAX}
\frac{\dd^2 Q}{\dd x^2}+(A_{\rm ax} \,\omega^2-V_{\mathrm{ax}})Q=0\,,
\end{equation}
where the tortoise coordinate $x$ is defined as
\begin{equation}
x_{,r} = 
\frac{\dd x}{\dd r}
= (A B)^{-1/2}\,\upd{,}
\label{eq:def_tortoise}
\end{equation}
$V_{\rm{ax}}$ is the potential for the axial perturbation while the function $A_{\rm ax}$ is given by 
\begin{align}
\label{eq:freq_coeffAX}
A_{\rm ax} &=\frac{A}{A - 2 \alpha  B A' \phi_0' f_0'} 
\{ 1-2 \alpha  B' \phi_0' f_0'
\nonumber \\
&\quad + 4 \alpha  B [{\phi_0'}^2 f_0'' + \phi_0''
   f_0'] \}\,,
\end{align}
(where primes on $A$, $B$ and $\phi_0$ refer to radial derivatives) or in the small coupling limit valid to $\mathcal{O}(\alpha^2)$
\be
\label{eq:freq_coeffAXsc}
A_{\rm ax} =1-\frac{\alpha^2\dfzs}{r^3}\left(16+\frac{16}{r}+\frac{32}{r^2}-\frac{256 }{r^3}\right)\,.
\ee
In the GR limit, $A_{\rm ax}$ reduces to unity  and Eq.~\eqref{eq:masterAX} reduces 
to the familiar Regge-Wheeler equation.
We may make yet another radial coordinate transformation:
\be
\label{eq:preikcoord}
\frac{\dd \tilde{x}}{\dd x}=\sqrt{A_{\rm ax}}\,,
\ee
after which  Eq.~\eqref{eq:masterAX} takes the form
\be
\label{eq:preikeqn}
\frac{\dd^2 Q}{\dd\tilde{x}^2}
+
p_{\rm ax}\frac{\dd Q}{\dd \tilde{x}}
+
\left(\omega^2-\tilde{V}_{\rm ax}\right)Q=0\,,
\ee
where we have defined the friction coefficient as $p_{\rm ax} = {(A_{\rm ax})_{,x}}/({2A_{\rm ax}^{3/2}})$
and the resulting effective potential $\tilde{V}_{\rm ax} = V_{\rm ax}/A_{\rm ax}$. As we will see later, the friction term makes no contribution to the QNM frequency in the eikonal approximation.
The expression for the potential $\tilde V_{\mathrm{ax}}(r)$ is rather lengthy and can be found in the supplemental Mathematica notebook~\cite{mathematica}. 

\subsection{Polar perturbations}
\label{sec:polar_pert}

The polar sector of the perturbations is somewhat 
more complicated as a result of the coupled 
tensorial and scalar perturbations.
The tensorial perturbations are written as~\cite{Blazquez-Salcedo:2016enn}
\begin{gather}
\label{eq:metric_pert_Y}
h_{ab}=\begin{pmatrix} A \bar H_0 && \bar H_1 && 0 && 0\\
\bar H_1 && \bar H_2/B && 0 && 0\\
0 && 0 && r^2\bar K && 0\\
0 && 0 && 0 && r^2\sin^2\theta \bar K \end{pmatrix} 
\, Y_{\ell m} \,.
\end{gather}
Once again, $\bar H_0$, $\bar H_1$ and $\bar K$ are functions of $(t,r)$ which we Fourier transform following Eq.~\eqref{eq:Fourier}.
The scalar field perturbation is decomposed in a simliar way as
\begin{equation}
\label{eq:scalar_pert_Y}
\delta \phi  = \frac{1}{\sqrt{2\pi}} \int \dd t \, 
\frac{\hat{\phi}(r)}{r} Y_{\ell m} \, 
e^{-i \omega t}\,.
\end{equation}

Inserting these expressions in the field equations, 
we arrive at a system of six coupled equations [arising from Eq.~\eqref{eq:gravity}] and one 
from Eq.~\eqref{eq:scalar}. 
(These equations for EdGB are shown in Ref.~\cite{Blazquez-Salcedo:2016enn}, Appendix B).
Using all seven equations, we can eliminate $\bar H_0$ and $\bar H_2$ so that the remaining first-order system of differential equations takes the form~\cite{Blazquez-Salcedo:2016enn}:
\begin{equation}
\label{eq:matrix_form}
\begin{pmatrix}
\bar H_1' \\ \bar K' \\ \hat{\phi}' \\ \hat{\phi}''
\end{pmatrix}+\begin{pmatrix}
V_{11} && V_{12}  && V_{13} && V_{14}\\
V_{21} && V_{22} && V_{23} && V_{24}\\
0&& 0 && 0 && -1\\
V_{41} && V_{42} && V_{43} && V_{44}
\end{pmatrix}\begin{pmatrix}
\bar H_1\\ \bar K \\ \hat{\phi} \\ \hat{\phi}'
\end{pmatrix}=\begin{pmatrix}
0 \\ 0\\0\\0
\end{pmatrix}.
\end{equation}

Following the original treatment by Zerilli~\cite{Zerilli:1970se}, the two first order gravitational perturbation equations may be rewritten as a single second order differential equation. By means of the field redefinitions
\begin{align}
\label{eq:field_redefine1}
\bar K(r) &= g(r)\hat{K}(r)+\hat{R}(r)\,,
\\
\label{eq:field_redefine2}
\bar H_1(r)  &= \omega \, (h(r)\hat{K}(r)+k(r)\hat{R}(r)) \,,
\end{align}
choosing $g$, $h$ and $k$ such that
\begin{align}
\label{eq:schro_form1}
\frac{\dd \hat{K}}{\dd x} &= \hat{R}\,, \\
\label{eq:schro_form2}
\frac{\dd \hat{R}}{\dd x} &= [A_0+A_2\omega^2]\hat{K}\,,
\end{align}
we obtain an inhomogeneous Schr\"odinger-type equation for $\hat{K}$; the non-GR source term of that equation depends on $\omega$, $\hat{\phi}$, and $\hat{\phi}'$. The functions $A_0(r)$ and $A_2(r)$ originate from the field redefinitions~\eqref{eq:field_redefine1} and~\upd{\eqref{eq:field_redefine2}}. 

The final distilled form of the polar perturbation equations is a system of two coupled wave-equations 
\begin{align}
&\frac{\dd^2\hat{K}}{\dd x^2}+p_{\rm pol} \frac{\dd\hat{K}}{\dd x}
+ ( A_{\mathrm{pol}}\, \omega^2-V_{\mathrm{pol}}) \hat{K} = a_0 \hat{\phi} + a_1 \frac{\dd \hat{\phi}}{\dd x}\,, 
\nonumber \\
\label{eq:final1} \\
\label{eq:final2}
&\frac{\dd^2\hat{\phi}}{\dd x^2}
+ (\omega^2-V_{\phi})\hat{\phi} = b_0 \hat{K}+b_1 \frac{\dd\hat{K}}{\dd x}\,.
\end{align}
Here \upd{$p_{\mathrm{pol}}$,~$A_{\mathrm{pol}}$,~$a_0$,~$a_1$,~$b_0$, and $b_1$} are functions of $r$ whose explicit forms in the small coupling approximation are given in Appendix~\ref{app:supplemental} and the supplemental Mathematica notebook~\cite{mathematica}. The potential $V_{\mathrm{pol}}$ for the gravitational perturbation equation is given, 
also in the small coupling approximation, as
\be
\label{eq:V_sGB_polar}
V_{\mathrm{pol}}(r) = V_{\rm Z}(r) + V_{2}(r)\,\alpha^2 \dfzs \,.
\ee 
Here, $V_{\rm Z}$ is the Zerilli potential~\cite{Zerilli:1970se}
\begin{align}
\label{eq:VZ}
V_{\rm Z} = \left( 1-\frac{2}{r} \right) 
\frac{2\Lambda^2 (\Lambda+1)r^3 
+ 6\Lambda^2 r^2 
+ 18\Lambda r 
+ 18}{r^3 (\Lambda r+3)^2}\,,
\end{align}
with 
\be
\label{eq:Lambda}
\Lambda = (\ell+2)(\ell-1) / 2 \,,
\ee
while $V_2$ and the scalar perturbation potential $V_\phi$ [appearing 
in Eq.~\eqref{eq:final2}] are given in Appendix~\ref{app:supplemental} and the supplemental Mathematica notebook~\cite{mathematica}.
Taking the GR limit ($\alpha \to 0$) removes all the right-hand side
coupling terms in Eqs.~\eqref{eq:final1} and~\eqref{eq:final2} and reduces the 
left-hand-sides
to the Zerilli and free 
scalar field wave equations \upd{respectively}.
Note that the system, Eqs.~\eqref{eq:final1}--\eqref{eq:final2}, does not belong to the general 
family of coupled equations studied in~\cite{Glampedakis:2019dqh,Silva:2019scu}. 

\section{Eikonal QNMs}
\label{sec:eikonal_qnm}

Having obtained the equations governing axial~\eqref{eq:masterAX}
and polar~\eqref{eq:final1}--\eqref{eq:final2} perturbations in scalar Gauss-Bonnet gravity, we now proceed to analyze their QNM
spectra in the eikonal limit using the methods developed in  the previous papers of the series~\cite{Glampedakis:2019dqh,Silva:2019scu}.

\subsection{Axial QNMs}
We start off with the axial sector which provides a simple 
setup to review these methods.
Here we use coordinates given by Eqs.~\eqref{eq:def_tortoise} and~\eqref{eq:preikcoord}, which may be expressed in the small coupling limit as
\begin{align}
\tilde{x}_{,r} &= \left( 1 - \frac{2}{r} \right)^{-1}
\left[
1 +
\frac{\alpha^2 \dfzs}{r}
\left(
\frac{49}{40}
+ \frac{39}{20 r}
- \frac{143}{30 r^2}
\right. \right.
\nonumber \\
&\left. \left.
\quad 
- \frac{218}{15 r^3}
- \frac{484}{15 r^4}
- \frac{272}{3 r^5}
\right)
\right]\,.
\label{eq:preikcoord_expanded}
\end{align}
The eikonal prescription is based on a phase-amplitude solution of the form
\be
    Q(\tilde{x}) = \mathcal{A}_{Q}(\tilde{x}) \, e^{i S(\tilde{x})/\epsilon}\,,
    \label{eq:q_eik_anstaz}
\ee
where $\epsilon$ is the eikonal bookkeeping parameter.
The eikonal limit corresponds to $\epsilon \ll 1$ and $\ell \gg 1$, while keeping the balance $\epsilon \ell = {\cal O}(1)$.
For later convenience, we decompose the potential into
\be
\tilde{V}_{\mathrm{ax}}=\ell(\ell+1) V_{\mathrm{ax1}}+V_{\mathrm{ax2}}\,,
\ee
where $V_{\mathrm{ax1}}$ and $V_{\mathrm{ax2}}$ are independent of both $\ell$ and $\omega$; thus only the former function can contribute to the QNM spectra in the eikonal limit.

\subsubsection{Leading\upd{-}order analysis}

Substituting the ansatz~\eqref{eq:q_eik_anstaz} into Eq.~\eqref{eq:preikeqn}, we find the following leading order eikonal equation:
\be
\label{eq:leadax}
     - \frac{\left(S_{,\tilde{x}}\right)^2}{\epsilon^2} +\omega^2 - \ell^2 V_{\rm ax1}=0\,.
\ee
The explicit expression for the effective potential $V_{\rm ax1}$ vanishes for arbitrarily large $|\tilde{x}|$ with a peak, located at a radial position denoted $r_{\rm m}$, where $V_{\rm ax1}'(r_{\rm m})=(V_{\rm ax1}')_{\rm m}=0$. The derivative of Eq.~\eqref{eq:leadax} evaluated at $r_{\rm m}$ yields
\be
\label{eq:peakdef}
\frac{2}{\epsilon^2}(S_{,\tilde{x}})_{\rm m}(S_{,\tilde{x}\tilde{x}})_{\rm m}=-\ell^2\left(\frac{\dd r}{\dd\tilde{x}}\right)_{\rm m}(V'_{\rm ax1})_{\rm m}=0,
\ee
showing the potential is extremum at the same location where $S_{,\tilde{x}}=0$ given $S_{,\tilde{x}\tilde{x}}\neq0$. A location of stationary phase $S$ follows from imposing purely ingoing and outgoing plane wave solutions as $|\tilde{x}|\gg 0$; that is, purely ingoing towards the horizon and purely outgoing at spatial infinity,
requiring a minimum $S_{,\tilde{x}}=0$ already determined by Eq.~\eqref{eq:peakdef}.
Hence, Eq.~\eqref{eq:leadax} at this peak yields
\be
\omega^2=\ell^2 \, 
(V_{\rm ax1})_{\rm m}\,,
\ee
given explicitly by 
\be
\label{eq:real0}
\omega^{(0)}_{R}=\ell \, \left[\frac{A-2 \alpha  B A' \phi_0' \dfz}{r \left(r - 4 \alpha 
   B \phi_0' \dfz\right)}\right]^{1/2}_{\rm{m}} \,,
\ee
where the labels denote that this is the leading order real part of the QNM modes.

The condition $V_{\text{ax1},\,r}=0$ becomes
\begin{align}
\label{eq:axpeak}
(r_{\rm m}-3)& 
- \alpha^2 \dfzs \left( \frac{147}{80}+\frac{155}{6r_{\rm m}^2}-\frac{98}{r_{\rm m}^3} \right.
\nonumber \\
&\left. -\frac{77}{5r_{\rm m}^4}-\frac{1408}{5r_{\rm m}^5}+\frac{984}{r_{\rm m}^6}\right)=0,
\end{align}
and may be solved for $r_{\rm m}$ by means of a small coupling expansion ansatz. Solving to second order we obtain,
\be
\label{eq:Vpeaksc}
r_{\rm m}=3+\frac{6577}{19440}\alpha^2 \dfzs \,,
\ee
where the first term represents the GR photon ring 
i.e., the radius of the unstable photon circular orbit.

Hence the leading-order real mode can be expressed as
\be
\label{eq:omegaR0sc}
\omega_R^{(0)}=\frac{\ell}{3 \sqrt{3}}\left(1- \frac{71987}{174960}\alpha ^2 \dfzs \right)\,,
\ee
where we can again identify the first term as the appropriate GR limit.

\subsubsection{Subleading\upd{-}order analysis}

Let us next derive the QNM frequency at the subleading eikonal order. 
The subleading order equation evaluated at the potential peak gives
\be
\label{eq:subleadax}
\frac{i(S_{,\tilde{x}\tilde{x}})_{\rm m}}{\epsilon}-\ell (V_{\mathrm{ax1}})_{\rm m}
+2\left(\omega^{(1)}_{R}+i\omega_{I}^{(1)}\right)\omega^{(0)}_{R} =0\,,
\ee
where we used 
\begin{equation}
\omega = \omega_R^{(0)} + \epsilon \left(\omega^{(1)}_{R}+i\omega_{I}^{(1)}\right) + \mathcal{O}(\epsilon^2)\,.
\end{equation}
Taking the real part of Eq.~\eqref{eq:subleadax} and using Eq.~\eqref{eq:real0}, we find the subleading eikonal correction to the real part of the axial QNM frequency as 
\be
\label{eq:real1}
\omega^{(1)}_{R} = \frac{1}{2}
\left [\,
\frac{A-2 \alpha  B A' \phi_0' \dfz}{r \left (\, r-4 \alpha 
   B \phi_0' \dfz \,\right )} \, \right ]^{1/2}_{\rm m}
   = \frac{\omega^{(0)}_{R}}{2\ell}.
\ee
We can combine this expression with the leading order result to obtain
\begin{align}
\omega_{R} &= \omega^{(0)}_{R} + \epsilon \, \omega^{(1)}_{R}\,, 
\nonumber \\
&= \left(  \ell+\frac{\epsilon }{2}\right) 
\left [\, \frac{A-2 \alpha  B A' \phi_0' \dfz}{r\left( r -4 \alpha 
 B \phi_0' \dfz \,\right)} \,\right]^{1/2}_{\rm m}\,.
\end{align}
\upd{Finally,} using Eq.~\eqref{eq:Vpeaksc} and the background solutions, we find
\be
\label{eq:omegarsc}
\omega_R=\frac{1}{3 \sqrt{3}}\left(\ell+\frac{\epsilon }{2}\right)\left(1- \frac{71987}{174960}\alpha ^2 \dfzs \right).
\ee

Let us now derive the imaginary part. To do so, we need
$(S_{,\tilde{x}\tilde{x}})_{\rm m}$ which can be solved for by doing a Taylor expansion of the leading-order equation~\eqref{eq:leadax} around
$\rpk{}$ with $\omega$ given by Eq.~\eqref{eq:real0}, followed by a derivative 
with respect to $\tilde{x}$. 
These steps result in
\be
\frac{S_{,\tilde{x}}^2}{\epsilon^2}\approx-\frac{\ell^2}{2}(V_{\mathrm{ax1}}'')_{\rm m} (\tilde{x}-\tilde{x}_{\rm m})^2\,.
\ee
A Taylor expansion of the left-hand-side term about the peak radius
leads to
\be
\label{eq:ddS_axial}
\frac{(S_{,\tilde{x}\tilde{x}})_{\rm m}}{\epsilon}=\frac{\ell}{\sqrt{2} \, x_{,r}}
|V_{\mathrm{ax1}}''|^{1/2}_{\rm m}\,.
\ee
Finally, substituting Eq.~\eqref{eq:ddS_axial} in~Eq.~\eqref{eq:subleadax} gives,
\be
\label{eq:imag}
\omega_{I} = \epsilon \, \omega_{I}^{(1)} =  -\frac{1}{2\, (\tilde x_{,r})_{\rm m}} \sqrt{\frac{| V_{\mathrm{ax1}}''|}{2V_{\mathrm{ax1}}}} \, \Bigg|_{\rm m}.
\ee
Using further Eq.~\eqref{eq:Vpeaksc}, we obtain
\begin{align}
\label{eq:omega_im_ax}
\omega_I &= -\frac{\epsilon }{2\sqrt{3}\, (\tilde{x}_{,r})_{\rm m}}\left(1-\frac{115771}{174960}\alpha^2 \dfzs \right)\,, 
\nn \\
         &= -\frac{\epsilon }{6\sqrt{3}}\left(1-\frac{121907}{174960}\alpha^2 \dfzs \right),
\end{align}
which also recovers the well-known GR limit.

\subsubsection{Comparison with geodesic correspondence}

In~\cite{Blazquez-Salcedo:2016enn}, approximate QNM frequencies for axial modes were computed from the null geodesic correspondence in the eikonal limit~\cite{Ferrari:1984zz,Cardoso:2008bp,Yang:2012he} and were compared with numerical results. 
We here compare our eikonal calculations with the geodesic correspondence results.

The geodesic correspondence allows one to compute QNM frequencies only from properties of the photon ring.
The complex QNM frequency under this correspondence is related to the metric functions as~\cite{Cardoso:2008bp}
\begin{align}
\label{eq:omega_geod}
\omega^{\mathrm{(geod)}} = \ell 
\frac{\sqrt{A(r_c)}}{r_c} - i \frac{r_c}{2\sqrt{2}} \sqrt{-\frac{1}{A(r_c)} \left( \frac{\dd^2}{\dd \tilde{x}^2} \frac{A}{r^2}\right)_{r=r_c}}\,,
\nonumber \\
\end{align}
where $r_c$ is the location of the photon ring determined from the equation
\begin{equation}
2 A(r_c) = r_c A'(r_c)\,.
\end{equation}
For scalar-Gauss-Bonnet gravity \upd{and} in the small coupling approximation, we can use Eq.~\eqref{eq:metric_A} and solve this equation for $r_c$ order by order in $\alpha$ to 
find
\begin{equation}
\label{eq:light_ring}
r_c = 3 +\frac{4219}{6480}\alpha ^2 f'_0{}^2\,,
\end{equation}
to second order in $\alpha$.

Let us first study the real part of the QNM frequency
\begin{equation}
\label{eq:omega_R_geod}
\omega_R^{\mathrm{(geod)}} = \ell \frac{\sqrt{A(r_c)}}{r_c}\,. 
\end{equation}
From Eqs.~\eqref{eq:metric_A} and~\eqref{eq:light_ring}, we find
\begin{equation}
\omega_R^{\mathrm{(geod)}} = \frac{\ell}{3 \sqrt{3}}\left(1- \frac{71987}{174960}\alpha ^2 \dfzs \right)\,.
\end{equation}
Notice that this is exactly the same as $\omega_R^{(0)}$ in Eq.~\eqref{eq:omegaR0sc} obtained from the eikonal calculation. At a first glance, this seems a bit surprising since Eq.~\eqref{eq:real0} contains the scalar field dependence whereas Eq.~\eqref{eq:omega_R_geod} does not, and the right hand side of the former equation is evaluated at $r_{\mathrm{m}}$ which is different from $r_c$.

The apparent difference in the real part of the QNM frequency in the two analyses mentioned above does not affect the final expression under the small coupling approximation for the following reason. First, let us look at the two terms in Eq.~\eqref{eq:real0} that involve the scalar field $\phi_0$. Given that these are already multiplied by $\alpha$ and $\phi_0 = \mathcal{O}(\alpha)$, we can replace $A' \to A'_\GR = \dd (1-2/r)/ \dd r = 2/r^2$, $B \to B_\GR = 1-2/r$, $\phi_0' \to \alpha \phi_1'$ and $f'(\phi_0) \to f_0'$ if we only work up to $\mathcal{O}(\alpha^2)$, where the subscript ``GR'' denotes the GR contribution and $\phi_1$ is the $\mathcal{O}(\alpha)$ piece in $\phi_0$ (with $\alpha$ being factored out). Replacing further $A \to A_\GR + \alpha^2 \delta A$ and $r_{\rm{m}} \to 3+ \alpha^2 \delta r_{\rm{m}}$ with $\delta A$ and $\delta r_{\rm{m}}$ being some generic functions that are independent of $\alpha$, we find
\begin{align}
\omega^{(0)}_{R} &= \ell \, \left [\, \frac{A_\GR + \alpha^2 \delta A-2 \alpha^2  B_\GR A_\GR' \phi_1' f'_0}{r \left(r - 4 \alpha^2 
   B_\GR \phi_1' f'_0\right)}\,\right ]^{1/2}_{r=3+\delta r_{\rm{m}}} \nonumber \\
   &\approx \frac{\ell}{3\sqrt{3}}\left(1  + \frac{3}{2} \alpha ^2 \delta A(3) \right)\,.
\end{align}
Notice that the GB correction only depends on $\delta A$ and are independent of $\phi_1$ and $\delta r_{\rm{m}}$. Also notice that we only need to evaluate $\delta A$ at the GR value for $r_{\rm{m}}$, namely $r=3$. Substituting in $\delta A(3) = - 71987 f_0'{}^2/262440$, we recover Eq.~\eqref{eq:omegaR0sc}. The above calculation proves analytically that the scalar field (and also $\delta r_{\rm{m}}$) dependence in Eq.~\eqref{eq:real0} cancels at $\mathcal{O}(\alpha^2)$, leading to the same expression for the real QNM frequency as in the geodesic correspondence.

Next, we study the imaginary part of the QNM frequency in the geodesic side of the correspondence. From Eq.~\eqref{eq:omega_geod}, together with Eqs.~\eqref{eq:metric_A},~\eqref{eq:metric_B}, and~\eqref{eq:light_ring}, we find to $\mathcal{O}(\alpha^2)$
\begin{align}
\omega_I^{\mathrm{(geod)}} &= -  \frac{r_c}{2\sqrt{2}} \sqrt{-\frac{1}{A(r_c)} \left( \frac{\dd^2}{\dd \tilde{x}^2} \frac{A}{r^2}\right)_{r=r_c}}\,, \nonumber \\
&\approx  -\frac{1}{6\sqrt{3}}\left(1-\frac{121907}{174960}\alpha^2 \dfzs \right)\,.
\end{align}
Once again, this is same as the eikonal result in Eq.~\eqref{eq:omega_im_ax}. 
In conclusion, our eikonal QNM calculation agree with those from the geodesic correspondence up to $\mathcal{O}(\alpha^2)$ for the axial modes.

\subsection{Polar QNMs}
\label{sec:polarQNMs}

Let us next study the eikonal QNM frequencies in the polar sector. The coupled wave equations describing polar QNMs are given in Eqs.~\eqref{eq:final1} and~\eqref{eq:final2} valid to $\mathcal{O}(\alpha^2)$.
As we did in the axial case, we start by introducing the eikonal ansatz,
\be
\hat{K} (x) = \mathcal{A}_{K}(x) \, e^{iS(x)/\epsilon}, 
\quad
\hat{\phi} (x) = \mathcal{A}_{\phi} (x) \, e^{i S(x)/\epsilon}\,.
\label{eq:eikonal_ansatz}
\ee
Note that both fields share the same phase function $S$ (this should not be confused with the 
previous axial phase function). As already pointed out, we assume an eikonal scaling 
$\ell = {\cal O} (\epsilon)$ which is appropriate for  standard ``Price" QNMs.
Similar to the axial case, the leading order frequency of these modes is $\omega_R$ while $\omega_I$ 
first appears at subleading order. 

On paper, the strategy for manipulating the wave equations should be simple: 
after using Eq.~\eqref{eq:eikonal_ansatz}, we solve the tensorial equation 
for $\mathcal{A}_{\phi}$ and then insert the result in the scalar equation. 
The outcome is an algebraic biquadratic equation for $\omega$ which is supposed 
to be solved at the peak radius $r=\rpk{}$ (once again, not to be confused with the peak location for the axial potential) of an effective potential similar to Eq.~(40) of~\cite{Glampedakis:2019dqh}. In the previous papers of this series~\cite{Glampedakis:2019dqh,Silva:2019scu} the $\epsilon \to 0$ limit  was applied to 
this equation (or equivalently to its solutions), resulting in eikonal expressions 
for $\omega$ (up to a specified order).
Taking the eikonal limit in the present analysis requires a  more subtle computation
due to the presence of a second small parameter in the system, the coupling constant 
$\alpha$. The polar calculation is essentially a biparametric expansion in $\epsilon \ll 1$ 
and $\alpha \ll 1$ and one has to make a prior decision as to whether $\alpha/\epsilon$ 
is supposed to be a small or a large parameter. This is a necessary step because, as we show below, 
taking the eikonal limit before expanding in $\alpha$ is \emph{not} equivalent to the same limits 
taken in the reverse order. 

The equation for $\omega$ can be symbolically written as,
\be
\omega^4 + F(\epsilon,\alpha, \ell,\rpk{},Q_{\rm m}) \, \omega^2 
+ G(\epsilon,\alpha,\ell,\rpk{},Q_{\rm m}) = 0\,, 
\label{EqomFull}
\ee
where $F$ and $G$ are rational functions of their arguments and
$Q = \{S^{\prime\prime}, \mathcal{A}_{K},\mathcal{A}_{K}^\prime,\mathcal{A}_{K}^{\prime\prime}
,\mathcal{A}_{\phi}^\prime,\mathcal{A}_{\phi}^{\prime\prime} \}$, where primes now represent $x $ derivatives.
In the double limit $\epsilon =\alpha =0$ this equation reduces to
\be
\left [\omega^2 - \frac{\ell^2}{\rpk{2}}\left( 1-\frac{2}{\rpk{}} \right)  \right]^2 = 0\,,
\ee
with the familiar GR double root $\omega_{\rm GR}^2 = \ell^2/27$ (with 
$\rpk{}=3$) for gravitational and scalar perturbations.

We now solve Eq.~\eqref{EqomFull} for nonvanishing values of $\epsilon, \alpha$ 
in combination with the ansatz for the peak location given by
\be
r_{\rm m} = 3 + \epsilon r_{01} + \alpha (r_{10} + \epsilon r_{11}) 
+ \alpha^2 ( r_{20} +  \epsilon r_{21} )\,.
\label{rmansatz2}
\ee
We can proceed following the same recipe as in the axial case finding that $r_{10} = r_{01} = 0$ and 
\begin{equation}
    r_{20} = - \frac{32}{27}\frac{\dfzs \ell \alpha^2}{\epsilon}\,.
\end{equation}
Unlike the others, this contribution to the peak location will be required in the $S''_{\rm m}$ calculation later.

Our first approach is that of ``eikonal limit before small-$\alpha$ expansion'' 
(which amounts to $\epsilon \ll \alpha$). We find the pair of roots,
\begin{align}
\omega^2_\pm &= \frac{\ell^2}{27} \left \{ 1 + \frac{\epsilon}{\ell} 
\left ( 1- \frac{27 i}{\ell} S^{\prime\prime}_{\rm m} \right ) \right.
\nn \\
&\quad \left. \pm \frac{8}{9} \dfz \alpha \left [ 1 + \frac{\epsilon}{\ell}
\left (1- \frac{4i}{\ell} S^{\prime\prime}_{\rm m} \right )  \right] \right.
\nn \\
&\quad \left. - \frac{\alpha^2}{3} \left [ \frac{67307}{58320}\dfzs{} 
 \mp \, \frac{320}{243} \dfz  \ddfz{} 
\right ] \right \} 
 +{\cal O} (\alpha^3, \epsilon^2, \alpha^2 \epsilon)\,.
\label{approach1_roots}
\end{align}
Among other things, these display a characteristic linear $\alpha$-dependence
at leading eikonal order. This is not too surprising given that both functions $F$ 
and $G$ in Eq.~\eqref{EqomFull} contain linear-$\alpha$ terms. 

The second approach is that of ``eikonal limit after small-$\alpha$ expansion'' (which amounts to $\alpha \ll \epsilon$).
Application of this algorithm [together with Eq.~\eqref{rmansatz2}] to the 
roots of Eq.~\eqref{EqomFull} leads to
\begin{align}
\omega_\pm^2 &= \frac{\ell^2}{27} \left \{\, 1 
+ \frac{\epsilon}{\ell} \left (1- 27 i\frac{S^{\prime\prime}_{\rm m}}{\ell} \right )
\right.
\nn \\
&\quad \left. 
\pm \frac{8\,\alpha^2  \dfzs{}}{27 \epsilon} 
\left [\,\frac{\ell^2}{\epsilon}+ 2 \left (\ell -4 i S_{\rm m}^{\prime\prime} \right )\,\right ]
\,\right\} 
+ {\cal O} (\alpha^3, \epsilon^2).
\label{approach2_roots}
\end{align}
This new pair of roots is clearly not the same as the one obtained earlier; the most striking 
difference is the absence of a linear-$\alpha$ leading-order correction and the unconventional 
eikonal scaling of the ${\cal O} (\alpha^2)$ piece. 
The linear correction first appears
at ${\cal O} (\epsilon^2 )$ and is therefore omitted in Eq.~\eqref{approach2_roots}. 
Although terms scaling as $\sim \epsilon^{-2}, \epsilon^{-1}$ are formally of 
leading eikonal order, the fact that they appear together with $\alpha^2$ effectively reduces 
their perturbative order and makes them smaller than the first GR term. This is equivalent to 
saying that $\alpha \ll \epsilon$, i.e. the opposite arrangement to that of the first approach. 
Another difference is the absence of $\ddfz$ terms in Eq.~\eqref{approach2_roots} relative
to Eq.~\eqref{approach1_roots}. This says that modulo a trivial rescaling of the coupling constant,
the second approach predict a `theory degeneracy' between shift-symmetric and dilatonic 
scalar-Gauss-Bonnet theories whose $\dfz$ expressions differ by a constant at most.
As we will see later, a $\ddfz$ dependence does exist, but only at higher eikonal orders.

Which of the these two non-equivalent approaches should we trust? The fact that we are looking 
for QNMs with a smooth GR limit suggests that $\alpha \ll \epsilon$ 
(i.e., the second approach) is the appropriate ordering of small parameters.
As we discuss below, this choice is also the one in agreement with the numerical QNM data of 
Ref.~\cite{Blazquez-Salcedo:2016enn}. Taking the square root of Eq.~\eqref{approach2_roots} we obtain 
the following solutions for the real and imaginary parts of $\omega$ up to subleading eikonal order: 
\begin{align}
\label{eq:omega_R_polar}
\omega_{R \pm} &= \frac{\ell}{3\sqrt{3}} \left [\, 1 + \frac{\epsilon}{2\ell} 
\pm \frac{4}{27} \frac{\alpha^2 \ell^{2} \dfzs{}}{\epsilon^2} 
\left (1 +  {{\frac{3\epsilon}{2\ell}}} \right ) \,\right ]\,, 
 \\
\label{eq:omega_I_polar}
\omega_{I\pm} &=  - \frac{3\sqrt{3}\epsilon}{2\ell }\left ( 1 \mp  \frac{44}{729} 
\frac{\alpha^2 \ell^2 \dfzs{}}{\epsilon^2} 
\right ) S_{\rm m}^{\prime\prime}\,.
\end{align}
The coupled character of the wave equations could in principle allow for exotic QNMs whose
leading order eikonal part is dominated by non-GR terms. Such modes would have no GR counterpart 
and would become trivial solutions $\omega \to 0$ in the $\alpha \to 0$ limit. 
We have not been able to find any QNMs with this property (nor they appear in the numerical analysis of Ref.~\cite{Blazquez-Salcedo:2016enn}).

The above expressions capture the modifications to the well-known GR eikonal expression 
$\omega_{s} = (\ell  + 1/2 + i/2) / (3 \sqrt{3})$ for the massless scalar and gravitational 
QNM frequencies of a Schwarzschild black hole. We see that the $\alpha^2$-corrections break the 
degeneracy between the eikonal QNMs of these two degrees of freedom. The splitting is symmetric, reminiscent of the Zeeman effect. (The same symmetric splitting of modes can also be caused 
by leading-order corrections in spin to the QNMs of a Schwarzschild black hole).

The missing ingredient to calculate $\omega_{I\pm}$ is an expression for $S''_{\rm m}$.
This result can be obtained through the same steps as done for the axial perturbations with the 
final result being:
\begin{align}
\label{eq:ddS}
    S''_{\rm m} = \frac{\ell}{27} \left(
    1 - \frac{560}{2187}\frac{\alpha^2 \ell \dfzs}{\epsilon} \right)\,.
\end{align}
Substituting this expression in Eq.~\eqref{eq:omega_I_polar} gives the final result 
for the imaginary part of $\omega$,
\begin{equation}
\label{eq:omega_I_pm}
    \omega_{I\pm} = - \frac{\epsilon}{6 \sqrt{3}} \left( 1 \pm \frac{44}{729} \frac{\alpha^2 \ell^2 \dfzs}{\epsilon^2} \right)\,.
\end{equation}

\section{Comparison against numerical results}
\label{sec:comparison}

Let us now compare the eikonal QNM frequencies with the ones found numerically in EdGB gravity in~\cite{Blazquez-Salcedo:2016enn}.
In this subsection, we use $f = \exp(\phi) / 4$ and thus $f'_0 = 1/4$.

\subsection{Axial Modes}

Let us begin with the axial modes. Our results for the QNM frequencies obtained in the previous section were expressed in terms 
of the bare Schwarzschild mass $M$. These can be rewritten in terms of the observable ADM mass 
$M_{*}$ using Eq.~\eqref{eq:ADM}. (The shift $M \to M_*$ introduces ${\cal O} (\alpha^2)$ corrections 
and therefore does not affect the normalized coupling constant.)
For the axial mode we find 
\begin{align}
 \omega_{R}&=\left(\ell+\frac{\epsilon }{2}\right)\frac{1}{3\sqrt{3}M_*}\left(1+\frac{4397}{21870}
  \alpha^2 \dfzs\right),
    \label{eq:RaxialADM}
    \\
   \omega_{I}&=-\frac{\epsilon }{6\sqrt{3}M_*}\left(1-\frac{1843}{21870}\alpha^2\dfzs\right).
    \label{eq:IaxialADM}
\end{align}

 \begin{figure}[t]
\includegraphics[width=8.5cm]{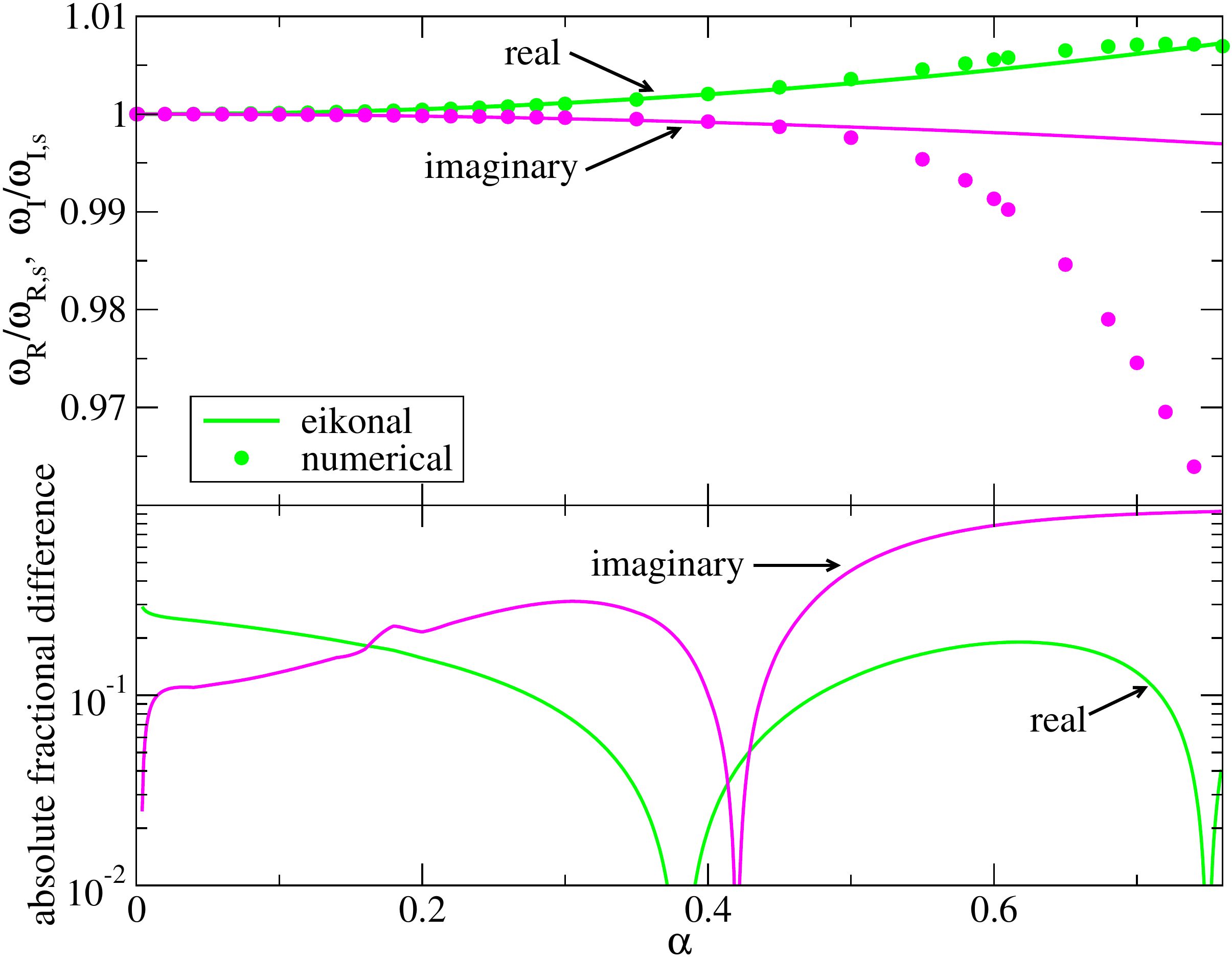}
\caption{\label{fig:axial_norm}
(Top) Real and imaginary axial QNM frequencies from eikonal [cf.~Eqs.~\eqref{eq:RaxialADM} and~\eqref{eq:IaxialADM}] and numerical results in EdGB gravity normalized by the corresponding Schwarzchild  case in GR with $\alpha=0$ ($\omega_{R,s}$ and $\omega_{I,s}$). 
(Bottom) Absolute fractional difference in Eq.~\eqref{eq:abs_frac_diff} between the eikonal and the numerical results.
}
\end{figure}

The top panel of Fig.~\ref{fig:axial_norm} presents the real and imaginary axial QNM frequencies normalized by the Schwarzschild case in GR ($\omega_s$) as a function of $\alpha$. We compare the analytic eikonal results with numerical ones.  The bottom panel shows the absolute fractional difference defined as
\begin{align}
\label{eq:abs_frac_diff}
\mathrm{(abs.~frac.~diff.)} &= \left| \frac{[(\omega/\omega_s)_{\mathrm{eik}} - 1] - [(\omega/\omega_s)_{\mathrm{num}} - 1]}{(\omega/\omega_s)_{\mathrm{num}} - 1} \right| \nonumber \\
&= \left|\frac{(\omega/\omega_s)_{\mathrm{eik}}  - (\omega/\omega_s)_{\mathrm{num}}}{(\omega/\omega_s)_{\mathrm{num}} - 1}\right|\,.
\end{align}
Namely, it measures the difference in the deviation of each curve from unity. For the real frequency, the eikonal calculation provides an accurate estimate within an error of $\sim 10\%$ (once the GR frequency has been corrected to the true value). For the imaginary frequency, the eikonal calculation is slightly worse and 
an error of $\sim 40\%$ for $\alpha \lesssim 0.5$. The eikonal calculation breaks down for large $\alpha$ since it is only valid to $\mathcal{O}(\alpha^2)$. We note that for non-normalized, raw frequencies, the real (imaginary) eikonal result has an error of 3\% (8\%) in GR.
\subsection{Polar Modes}
\label{sec:result_polar}

Let us next look at the polar modes. We present real and imaginary frequency results in turn.

\subsubsection{Real Frequency}

We begin by keeping only the leading eikonal corrections. When $\epsilon \ll \alpha$, the real QNM frequency can be computed from Eq.~\eqref{approach1_roots}, 
which is given in terms of the 
Schwarzschild mass $M$ (which has been set to 1). When converting this to the ADM mass $M_*$, one finds
\begin{eqnarray}
\label{eq:omega_R_polar_eik_epsilon}
\omega_{R\pm}^{(\epsilon \ll \alpha)} &=& \frac{\ell}{3
 \sqrt{3}M_*} \left(1 \pm \frac{4}{9} \alpha \dfz \right. \nn \\
 && \left.
 + \frac{112459
   \dfz\pm 76800 f_0''}{349920} \alpha^2 \dfz\right)\,.
\end{eqnarray}

On the other hand, when $\alpha \ll \epsilon$, the real part of the QNM frequency to leading eikonal order and leading scalar-Gauss-Bonnet correction is given by Eq.~\eqref{eq:omega_R_polar}.
The scalar-Gauss-Bonnet correction to the mass enters at $\mathcal{O}(\epsilon^0)$ which is of higher order than the $\mathcal{O}(\epsilon^{-2})$ above and thus can be neglected to leading order:
\begin{equation}
\label{eq:omega_R_polar_eik_alpha}
\omega_{R \pm}^{(\alpha \ll \epsilon)} = \frac{\ell}{3\sqrt{3}M_*} \left (\, 1  
\pm \frac{4}{27} \frac{\alpha^2 \ell^2 \dfzs{}}{\epsilon^2} \,\right )\,.
\end{equation}

\begin{figure}[t]

\includegraphics[width=8.5cm]{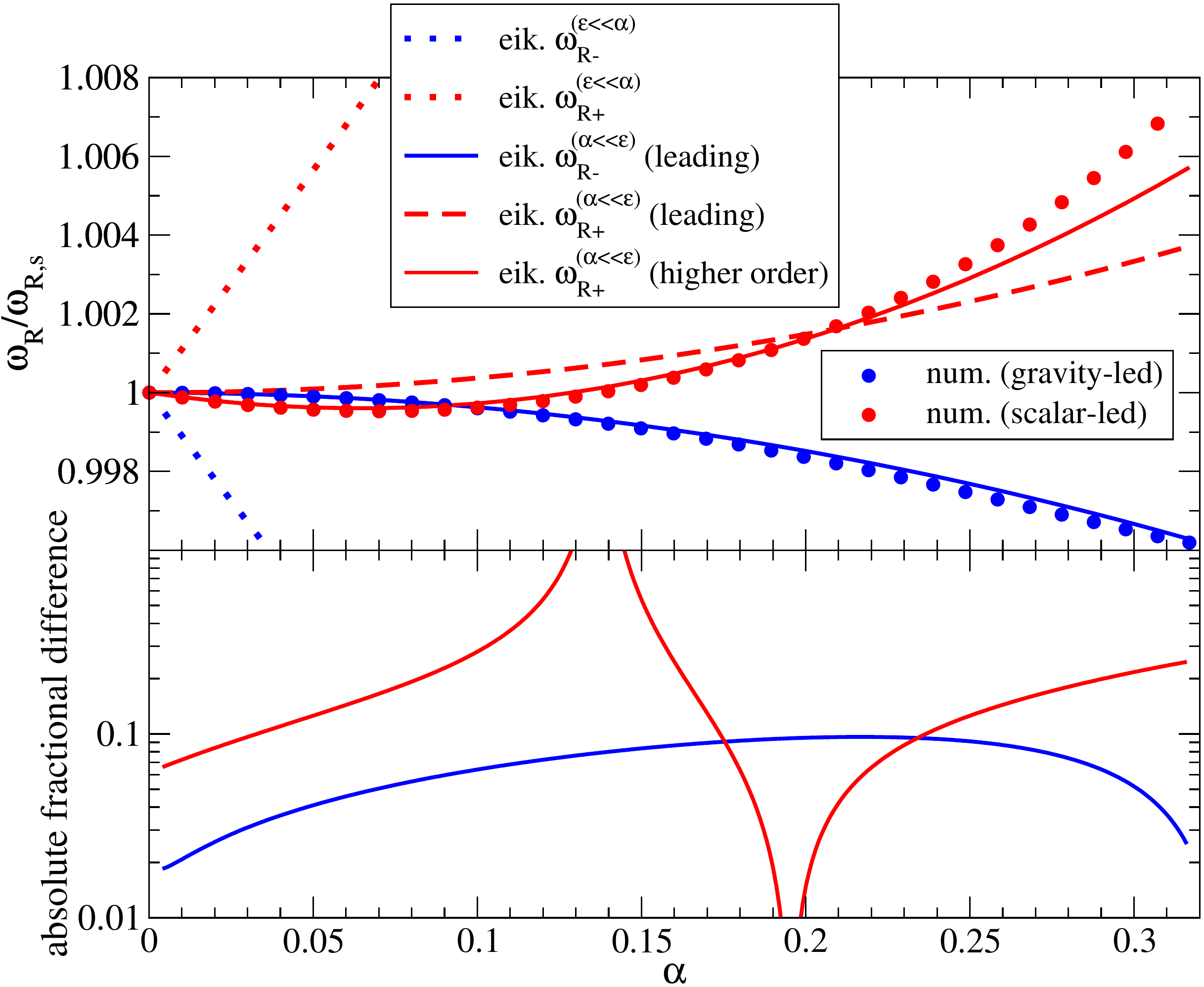}

\caption{\label{fig:polar_ratio_leading}
(Top) Normalized real QNM frequencies for the $\ell =2$ polar modes in EdGB gravity. We compare the leading eikonal calculations [cf.~Eqs.~\eqref{eq:omega_R_polar_eik_epsilon} and~\eqref{eq:omega_R_polar_eik_alpha}] and those with higher eikonal contribution for the + mode [cf.~Eq.~\eqref{eq:NNL}] with numerical ones. (Bottom) Absolute fractional difference in Eq.~\eqref{eq:abs_frac_diff} between eikonal and numerical calculations. The large difference at $\alpha \sim 0.13$ is an artifact of the numerical value crossing $\omega_R/\omega_{R,s} = 1$.
}
\end{figure} 

The top panel of Fig.~\ref{fig:polar_ratio_leading} presents the comparison between the above eikonal calculations and numerical results found in~\cite{Blazquez-Salcedo:2016enn}. Notice that the gravitational modes found numerically agree well with the eikonal ``negative'' mode within the assumption $\alpha \ll \epsilon$. On the other hand, the eikonal ``positive'' mode with $\alpha \ll \epsilon$ does not agree well with the numerical scalar-led mode. Furthermore, the eikonal calculations with $\epsilon \ll \alpha$ deviate significantly from the numerical results. This is because of a relatively large numerical coefficient at $\mathcal{O}(\alpha)$ in Eq.~\eqref{eq:omega_R_polar_eik_epsilon}.

How can we make the eikonal ``positive'' mode to agree better with the numerical scalar-led mode? The numerical result has a minimum at $\alpha \sim 0.08$ which cannot be realized by the eikonal leading result in Eq.~\eqref{eq:omega_R_polar_eik_alpha} since it is monotonically increasing in $\alpha$. To overcome this, one can take into account higher order contributions in the eikonal expansion. We found that $\mathcal{O}(\alpha)$ contribution enters at $\mathcal{O}(\epsilon^2)$ in $\omega_+^2$ in Eq.~\eqref{approach2_roots}. Keeping only the real contribution, we find\footnote{The imaginary part of $\omega^2$ contributes to $\omega_{R+}$ at $\mathcal{O}(\alpha^2 \epsilon^0)$ which we do not consider for simplicity.}
\begin{align}
\label{eq:omega_R_p_2}
\upd{\left(\omega_{R+}^{(\alpha \ll \epsilon)} \right)^{2}}
&= \frac{1}{27}\left [\,\ell^2 + \epsilon \ell + \frac{2}{3}\epsilon^2 -\frac{16}{27} \alpha\epsilon^2 \ddfz \right. \nn \\
&\quad \left.  + \frac{8}{27} \frac{\alpha^2}{\epsilon^2}\ell^2\left(\ell^2+2\epsilon \ell + \frac{4}{3}\epsilon^2\right) \dfzs{}
\,\right ]\,,
\nonumber \\
\end{align}
where we neglected a term proportional to $ \mathcal{A}''_K$ as such terms are unknown within the eikonal framework. Since we found there are no real corrections at $\mathcal{O}(\alpha \epsilon^3)$ while the one at $\mathcal{O}(\alpha \epsilon^4)$ is proportional to $\mathcal{A}''_K$ that we neglect, the above expression corresponds to keeping up to next-to-next-to-leading eikonal contributions at \emph{each order in $\alpha$}. We found that the contribution of the term at $\mathcal{O}(\alpha^0 \epsilon^2)$ is negligible and thus we do not consider it from here on. We further convert the mass to the ADM mass $M_*$, take a square root, expand about $\alpha =0$ and keep up to $\mathcal{O}(\alpha^2)$ to find\footnote{We ignored a term at $\mathcal{O}(\alpha^2 \epsilon^4)$ that is unimportant.} 
\begin{eqnarray}
\label{eq:NNL}
\omega_{R+}^{(\alpha \ll \epsilon)} 
&=& \frac{\sqrt{\ell  (\ell +\epsilon )}}{3 \sqrt{3}M_*} \left\{ 1 - \frac{8\alpha  \epsilon ^2 }{27\ell  (\ell
   +\epsilon )}\ddfz  +\frac{4}{27}\frac{\alpha^2  \dfzs}{\epsilon^2(\ell +\epsilon)}  \right. \nn \\
   &&  \times \left. \left[\ell ^2 (\ell +2
   \epsilon )+ \frac{4}{3}\ell \epsilon^2  + \frac{1323}{320} \epsilon^2 (\ell +\epsilon) \right]  \right\}\,. \nonumber \\
\end{eqnarray}
Notice that the frequency now has a $f_0''$ dependence that was absent in the expression to leading eikonal order.
We present this result in the top panel of Fig.~\ref{fig:polar_ratio_leading}. Notice that the agreement with the numerical result has been improved. 

The bottom panel of Fig.~\ref{fig:polar_ratio_leading} presents the absolute fractional difference of selected eikonal estimate from the numerical values. For the gravity-led mode, the numerical result is recovered with an error smaller than $10\%$. For the scalar-led mode, the eikonal result including higher order contribution also  reproduces the numerical result with an error of $20\%$ or smaller in the most range of $\alpha < 0.3$. The eikonal result becomes less accurate for larger $\alpha$ as it is obtained within the small coupling approximation. An apparent large deviation around $\alpha \sim 0.13$ is an artifact of the numerical value crossing  $\omega_R/\omega_{R,s} = 1$.

\subsubsection{Imaginary Frequency}

We now compare the imaginary part of the eikonal polar frequency with numerical results. Similar to the real frequency case, $\omega_{I}$ with $\epsilon \ll \alpha$ does not reproduce the numerical data, so we focus on $\alpha \ll \epsilon$ given in Eq.~\eqref{eq:omega_I_pm}. Since the correction to the ADM mass is of higher eikonal order than the one in Eq.~\eqref{eq:omega_I_pm} and can be neglected, we can simply multiply the expression in Eq.~\eqref{eq:omega_I_pm} by $1/M_*$ to find the expression in the ADM mass.

\begin{figure}[t]
\includegraphics[width=8.5cm]{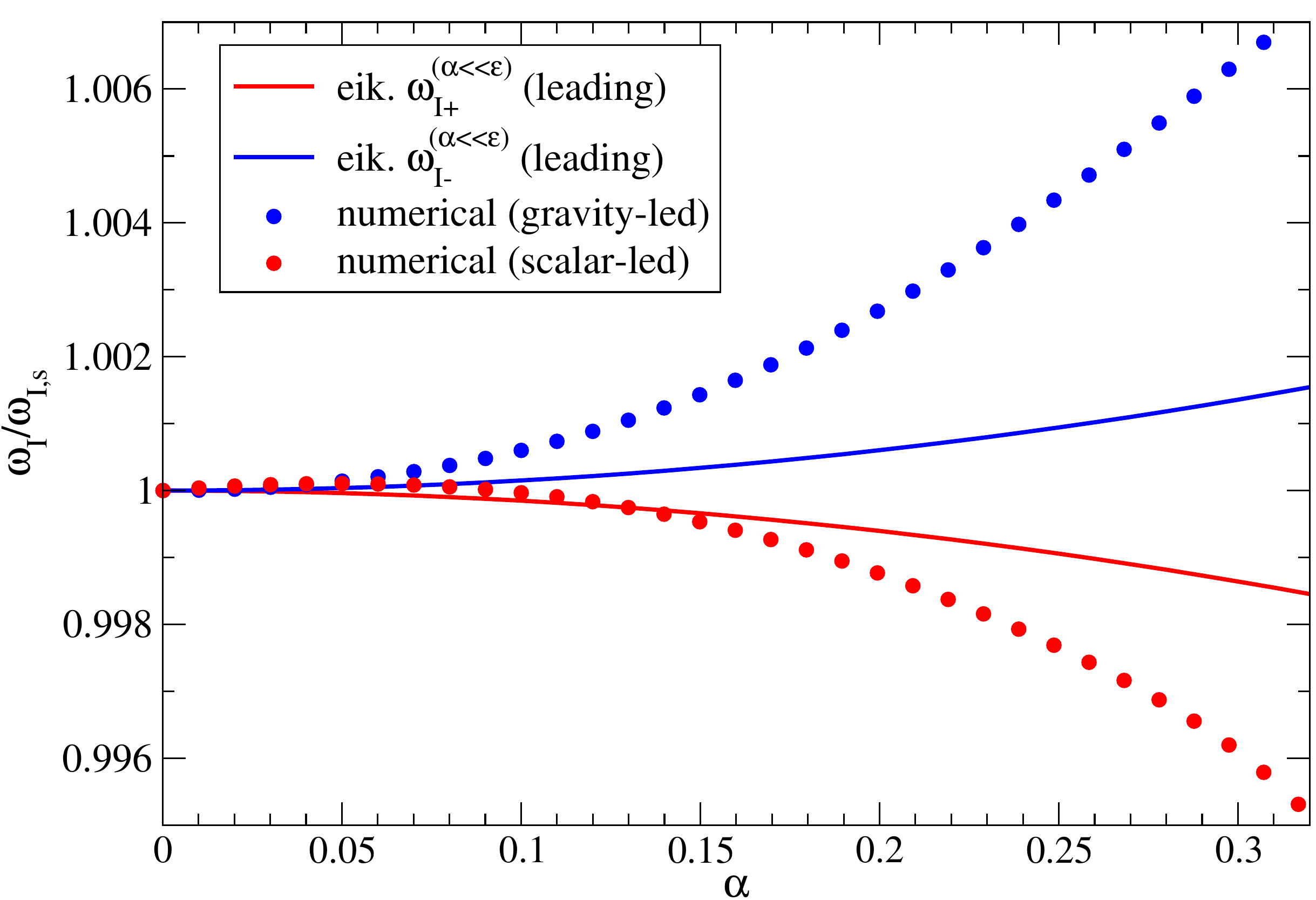}
\caption{\label{fig:Norm_Pol_Im} 
Normalized imaginary polar QNM frequencies for the $+$ and $-$ modes for $\ell=2$ in EdGB gravity. We compare the leading eikonal results given by Eq.~\eqref{eq:omega_I_pm} with the numerical ones.
}
\end{figure}

Figure~\ref{fig:Norm_Pol_Im} compares the analytic eikonal results to the numerical ones. The + ($-$) mode monotonically decreases (increases) in terms of $\alpha$, which is similar to the scalar-led (gravity-led) modes. However, the agreement between the two is not as good as the real frequency case. We have also tried including next-to-leading eikonal contributions but they did not improve the analytic results much. Unlike the real frequency case, the contribution at $\mathcal{O}(\alpha)$ is proportional to $\mathcal{A}''_{\phi}$ which is unknown in the eikonal framework. This may be one of the reasons why the eikonal results are less accurate for the imaginary frequency than the real one.

\subsection{Summary of Eikonal QNM frequencies}
\label{sec:summary}

We summarize here the final expressions for the eikonal QNM frequencies.

\subsubsection{Axial}

For axial modes, the real and imaginary eikonal frequencies are given by Eqs.~\eqref{eq:RaxialADM} and~\eqref{eq:IaxialADM}:
\begin{align}
 \omega_{R}&=\left(\ell+\frac{\epsilon }{2}\right)\frac{1}{3\sqrt{3}M_* }\left(1+\frac{4397}{21870}
  \alpha^2 \dfzs\right), \\
\omega_{I}&=-\frac{\epsilon }{6\sqrt{3}M_* }\left(1-\frac{1843}{21870}\alpha^2\dfzs\right).
\end{align}

\subsubsection{Polar}

For the polar modes \upd{and in the $\alpha \ll \epsilon$ limit}, the real eikonal frequencies for the gravity- and scalar-led modes are given respectively by Eqs.~\eqref{eq:omega_R_polar_eik_alpha} and~\eqref{eq:NNL}:
\begin{eqnarray}
\omega_{R -} &=& \frac{\ell}{3\sqrt{3}M_*} \left (\, 1  
- \frac{4}{27} \frac{\alpha^2 \ell^4 \dfzs{}}{\epsilon^2} \,\right )\,, \\
\omega_{R+}
&=& \frac{\sqrt{\ell  (\ell +\epsilon )}}{3 \sqrt{3}M_*} \left\{ 1 - \frac{8\alpha  \epsilon ^2 }{27\ell  (\ell
   +\epsilon )}\ddfz  +\frac{4}{27}\frac{\alpha^2  \dfzs}{\epsilon^2(\ell +\epsilon)}  \right. \nn \\
   &&  \times \left. \left[\ell ^2 (\ell +2
   \epsilon )+ \frac{4}{3}\ell \epsilon^2  + \frac{1323}{320} \epsilon^2 (\ell +\epsilon) \right]  \right\}\,. \nonumber \\ 
\end{eqnarray}
For the imaginary part, the leading eikonal result is given in Eq.~\eqref{eq:omega_I_pm} multiplied by $1/M_*$:
\begin{equation}
    \omega_{I\pm} = - \frac{\epsilon}{6 \sqrt{3}M_*} \left( 1 \pm \frac{44}{729} \frac{\alpha^2 \ell^2 \ddfz}{\epsilon^2} \right)\,.
\end{equation}

\section{Conclusions}
\label{sec:conclusions}

As a follow-up to our previous work~\cite{Glampedakis:2019dqh,Silva:2019scu}, in this paper we have studied the perturbations of nonrotating black holes in 
scalar Gauss-Bonnet gravity within a small-coupling approximation and have calculated the fundamental QNM frequencies in the eikonal/geometric optics approximation.

We first showed how the initial -- and rather complicated -- 
coupled perturbation equations (obtained in~\cite{Blazquez-Salcedo:2016enn}) can be reduced to a single modified Regge-Wheeler equation for the axial modes and a system of two coupled equations for the polar modes.
We subsequently applied the eikonal toolkit to these equations and analytically calculated the scalar-Gauss-Bonnet gravity 
modifications to the GR eikonal QNMs.
Among other things, this analysis allowed us to identify a key conceptual issue, namely, the correct ordering of the two underlying approximations
(small coupling limit and eikonal approximaiton); as a result we have found 
QNMs that are in good agreement with previous numerical calculations 
(with the exception of the polar mode's imaginary part).
The lessons learned from this study should be equally applicable to any
other theory that deviates perturbatively from GR. Such theories are ubiquitous, 
for instance, in effective field theory inspired extensions 
to GR (see e.g.~\cite{Endlich:2017tqa,Cardoso:2018ptl,Cano:2019ore}).

The present work can be extended in a number of ways. 
A simple generalization would be to repeat the calculation performed here using 
a higher $\alpha$-order black hole background; this would \upd{globally} improve 
the agreement between the eikonal formulae and the numerical results of Ref.~\cite{Blazquez-Salcedo:2016enn}.
A more sophisticated approach would be to abandon altogether the small-coupling approximation and work directly with a numerically determined black hole background.
A nonperturbative calculation along these lines would bring closer the
eikonal and numerical QNM results across the entire range of $\alpha$.
More importantly,
this calculation may
allow the eikonal study of QNMs of spontaneously scalarized Gauss-Bonnet 
black holes~\cite{Blazquez-Salcedo:2020rhf,Blazquez-Salcedo:2020caw} 
or perhaps even shed some more light on the mechanism of spontaneous scalarization
itself. It is also of interest to investigate the implications of losing hyperbolicity in Eqs.~\eqref{eq:schro_form2}--\eqref{eq:final2}, for the case of spontaneous scalarization as found in~\cite{Blazquez-Salcedo:2018,Blazquez-Salcedo:2020rhf,Blazquez-Salcedo:2020caw} and the existence of a second branch of modes that appears for larger Gauss-Bonnet and dilaton couplings, as found in~\cite{Blazquez_Salcedo:2017}. This should be studied in future work.

Another particularly important direction is to extend our calculation for rotating black hole spacetimes.
A first step in this direction has been taken in~\cite{Silva:2019scu} which worked perturbatively, to leading order in spin, on a parametrized pair of coupled  wave equations.
The strategy used in~\cite{Silva:2019scu} could be applied to the already analytically known slowly-rotating black hole 
solutions in scalar-Gauss-Bonnet gravity~\cite{Pani:2011gy,Maselli:2015tta} for which the perturbation equations could 
be obtained using the methods of~\cite{Kojima:1992ie,Kojima:1993ApJ,Kojima:PTP1993,Pani:2012vp,Pani:2012bp,Pani:2013pma},
and the QNM spectra calculated in~\cite{Pierini:2021jxd}.

Finally, from a conceptual point of view, it would be interesting to 
explore whether the geometrical optics--null geodesic correspondence, which 
was established here for the tensorial axial perturbations (described by a 
single wave equation) can be generalised to systems of coupled wave equations.

\acknowledgments
We thank Helvi~Witek for discussions and also Caio~F.~B.~Macedo for sharing unpublished numerical data used in this work.
K.Y. acknowledges support from NSF Award PHY-1806776, NASA Grant 80NSSC20K0523, a Sloan Foundation Research Fellowship, the Owens Family Foundation~\upd{and JSPS KAKENHI Grants No.~JP17H06358}.
K.Y. \upd{and K.G} would like to also acknowledge support by the COST Action GWverse CA16104.
H.O.S acknowledges support by the NSF Grant No.~PHY-1607130 and
NASA Grants No.~NNX16AB98G and No.~80NSSC17M0041.

\appendix

\bw
\section{Supplemental Expressions}
\label{app:supplemental}

In this Appendix, we show the explicit form of some of the functions introduced in the perturbations equations in the main text. 
All the expressions presented here, together with the axial perturbation potential $\tilde V_{\mathrm{ax}}$ in Eq.~\eqref{eq:preikeqn} and the GR polar (Zerilli) potential $V_{\rm Z}$ in Eq.~\eqref{eq:VZ}, are given in the supplemental Mathematica notebook~\cite{mathematica}.

First,  the function $c$ in Eq.~\eqref{eq:field_redefAX} for axial perturbations is given by
\begin{align}
c = k_1 \frac{\sqrt{B}}{r}\frac{A-2 \alpha  B A' \phi' f'(\phi)}{
  \sqrt{-2\alpha A B' \phi' f'(\phi) -4 \alpha  B\left[{\phi'}^2 f''(\phi) +\phi'' f'(\phi) \right]+1}}\,,
 \nonumber \\
\end{align}

with some arbitrary constant $k_1$.

Next, we present the functions appearing in the coupled gravito-scalar perturbation equations. 
The functions $p_{\rm pol}$, $A_{\mathrm{pol}}$, $a_0$ and $a_1$ in the gravitational perturbation equation~\eqref{eq:final1} are given by
\allowdisplaybreaks
\begin{align}
p_{\rm pol} &= 32\alpha ^2 \dfzs \frac{(r-2)^3 \left[96 + 3 (8 \Lambda -1) r + \Lambda r^2 + \Lambda  r^3\right]}{r^{10} (3 + \Lambda  r)^2}\,, \\
A_{\mathrm{pol}} &= 1-16 \alpha^2 \dfzs \frac{ (r-2) \left[32 - 24 r + 24 r^2 + (8 \Lambda + 11) r^3+3 (\Lambda + 1) r^4 + \Lambda  r^5\right]}{r^8 (3 + \Lambda r)}\,, \\
a_0 &= \frac{2\alpha \dfz}{r^8 (3 + \Lambda  r)^2}\bigg\{ 16 r^4 \omega ^2 (1-r) (3 + \Lambda  r) + (2 - r) [96 + 128 \Lambda  r + 8 [4 (\Lambda -3) \Lambda -21] r^2 \nonumber \\
&\quad -16 (2 \Lambda ^2+\Lambda -6) r^3+24
   \Lambda  (\Lambda +2) r^4 + 3 r^5+2 \Lambda  r^6]\bigg\}\,, 
\\
a_1 &= -16\alpha \dfz \frac{ (r - 2) 
\left[15 + (7 \Lambda -6) r - 3 \Lambda  r^2\right]}{r^5 (3 + \Lambda  r)^2}\,,
\end{align}
where $\Lambda = (\ell + 2) (\ell - 1) / 2$.
The functions in the scalar perturbation equation~\eqref{eq:final2} are given by
\allowdisplaybreaks
\begin{align}
\label{eq:field_coupling0}
b_0 &= \alpha  \left\{-\frac{4 \omega ^2 \dfz (r-2) (4 +r)}{r^4}+\frac{1}{r^8 (3 +\Lambda  r)^2}4 \dfz 
(r-2 ) \left[576 +18 (8 \Lambda -17)  r \right. \right. \nonumber \\
&\quad \left. \left. -3
   [22 \Lambda  (2 \Lambda +3)-3]  r^2+\Lambda  [9-2 \Lambda  (40 \Lambda +37)]  r^3+\Lambda ^2 [5-2 \Lambda  (6 \Lambda +5)] 
   r^4+\Lambda ^2 (\Lambda +1) r^5\right]\right\}  \nonumber \\
&\quad -\alpha ^2 \dfz \ddfz (r-2) \left\{\frac{\omega ^2 }{15  r^7} \left(800 +896  r+876  r^2+292  r^3+73 r^4\right) \right. \nonumber \\
&\quad\left.  -\frac{1}{15  r^{11} (3 +\Lambda  r)^2} \left[89280
   +288 (70 \Lambda +59)  r-24 \left(920 \Lambda ^2+732 \Lambda -1299\right)  r^2 \right. \right. \nonumber \\
&\quad\left.\left.  -4 [2 \Lambda  (8 \Lambda  (205 \Lambda
   +454)+963)+9063]  r^3-2 [2 \Lambda  (2 \Lambda  (4 \Lambda  (60 \Lambda +361)+3145)+5823)-657]  r^4 \right. \right. \nonumber \\
&\quad\left.\left.  +\left[657-2 \Lambda  \left(720
   \Lambda ^3+5468 \Lambda ^2+4310 \Lambda -657\right)\right] r^5+\Lambda  \left[657-2 \Lambda  \left(720 \Lambda ^2+574 \Lambda
   -365\right)\right]  r^6 \right. \right. \nonumber \\
&\quad\left.\left.  +73 \Lambda ^2 (2 \Lambda +5)  r^7+73 \Lambda ^2 (\Lambda +1) r^8\right]\right\}\,, \\
b_1 &= \frac{4\alpha \dfz (r-2) \left[96 +3 (8 \Lambda -1)  r+\Lambda   r^2+\Lambda  r^3\right]}{r^6 (3+\Lambda  r)} \nonumber \\
&\quad+\frac{\alpha ^2 \dfz \ddfz}{15  r^9 (3 +\Lambda  r)} (r-2) \left[14880+96
   (35 \Lambda +107)  r+4 (608 \Lambda +2583)  r^2 \right. \nonumber \\
&\quad \left. +4 (647 \Lambda -219)  r^3-219  r^4+73 \Lambda   r^5+73 \Lambda  r^6\right]\,.
\end{align}
The scalar potential in Eq.~\eqref{eq:final2} is only needed to ${\cal O}(\alpha)$ since $\phi_1(r)$ 
is already ${\cal O}(\alpha)$:
\begin{align}
\label{eq:scalar_potential}
V_\phi=\frac{2}{r^4} (r-2 ) (1+\Lambda  r+r)-\alpha f_0'' \frac{48  (r-2)}{r^7}\,.
\end{align}
Finally, the scalar-Gauss-Bonnet correction to the potential for gravitational perturbation $V_2$ in Eq.~\eqref{eq:V_sGB_polar} is given by
\begin{align}
V_2 &=-\frac{ (r-2) }{120  r^{12} (3 +\Lambda  r)^3}\bigg\{8847360+276480 (8 \Lambda -49) r-23040 [4 \Lambda  (22 \Lambda +57)-339]  r^2 \nonumber \\
&\quad   -192 [10
   \Lambda  (16 \Lambda  (40 \Lambda -29)-3999)+10953]  r^3-96 [4 \Lambda  (5 \Lambda  (16 \Lambda  (6 \Lambda -35)-1619)+10131)+2241] r^4  \nonumber \\
&\quad +24 [2 \Lambda  (40 \Lambda  (\Lambda  (96 \Lambda -1)-1327)-4641)+7353]  r^5 
-4 [4 \Lambda  (\Lambda  (8 \Lambda  (1135 \Lambda
   +4003)+4299)-9486)-8019]  r^6\nonumber \\
&\quad +2 [4 \Lambda  (2 \Lambda  (\Lambda  (68 \Lambda +485)+3744)+12501)+7209]  r^7 \nonumber \\
&\quad +[8 \Lambda  (\Lambda  (22
   \Lambda  (17 \Lambda +158)+10839)+2403)-3969]  r^8+\Lambda  [8 \Lambda  (\Lambda  (977 \Lambda +3143)+1332)-3969]  r^9 \nonumber \\
&\quad +9 \Lambda ^2 [4
   \Lambda  (23 \Lambda +72)-49]  r^{10} + 147 \Lambda ^2 [\Lambda  (2 \Lambda +5)+6] r^{11}\bigg\}\,.
\end{align}
\ew


\bibliography{ref}

\end{document}